\documentclass[journal]{IEEEtran}
\usepackage{amsmath,epsfig}
\usepackage{graphicx}
\usepackage{wrapfig}
\usepackage[caption=false]{subfig}
\usepackage{float}
\usepackage[export]{adjustbox}
\usepackage{hyperref}
\usepackage{tikz}
\usepackage{cite}
\usepackage{standalone}
\usepackage{pgfplots}
\usepackage{afterpage}
\usetikzlibrary{calc}

\usepackage{xcolor}

\pgfplotsset{compat=1.14}

\begin{document}
\thispagestyle{empty}
\vspace{100pt}
\begin{adjustbox}{valign=b,minipage=\textwidth}
\noindent © 2019 IEEE. Personal use of this material is permitted. Permission from IEEE must be obtained for all other uses, in any current or future media, including reprinting/republishing this material for advertising or promotional purposes, creating new collective works, for resale or redistribution to servers or lists, or reuse of any copyrighted component of this work in other works.
\end{adjustbox}

\newlength\e
\newlength\ee

\setcounter{page}{0}
\title{Short-term prediction of Electricity Outages Caused by Convective Storms}

\author{Roope Tervo,
        Joonas Karjalainen,
        Alexander Jung%
        \thanks{Manuscript received February 15, 2019; revised May 3, 2019 and May 31, 2019; accepted June 5, 2019. (Corresponding author: Roope Tervo.)}
        \IEEEcompsocitemizethanks{
        \IEEEcompsocthanksitem Roope Tervo is with Finnish Meteorological Institute, Observing and information systems centre, B.O. 503, 00101 Helsinki, Finland (e-mail: \protect roope.tervo@fmi.fi)
        \IEEEcompsocthanksitem Joonas Karjalainen is with Finnish Meteorological Institute, Observing and information systems centre, B.O. 503, 00101 Helsinki, Finland (e-mail: \protect joonas.karjalainen@fmi.fi)
        \IEEEcompsocthanksitem Alexander Jung is with Aalto University, Dept of Computer Science, B.O. 11000, 00076 Aalto, Finland
        \IEEEcompsocthanksitem Digital Object Identifier 10.1109/TGRS.2019.2921809}
        }

\maketitle

\begin{abstract}
Prediction of power outages caused by convective storms which are highly localised in space and time is of crucial importance to power grid operators. We propose a new machine learning approach to predict the damage caused by storms. This approach hinges identifying and tracking of storm cells using weather radar images on the application of machine learning techniques. Overall prediction process consists of identifying storm cells from CAPPI weather radar images by contouring them with a solid 35 dBZ threshold, predicting a track of storm cells and classifying them based on their damage potential to power grid operators. Tracked storm cells are then classified by combining data obtained from weather radar, ground weather observations and lightning detectors. We compare random forest classifiers and deep neural networks as alternative methods to classify storm cells. The main challenge is that the training data are heavily imbalanced as extreme weather events are rare.
\end{abstract}


\IEEEpeerreviewmaketitle

\section{Introduction}
\label{sec:intro}
A key problem faced by Finnish power grid operators is the prediction of damages caused by extreme weather events such as convective storms such as thunders which occur often in Finland during summer time \cite{Review2005Occurrence01}. These thunderstorms are typically geographically highly localised (50 km$^2$) and have a short duration (less than 30 minutes) \cite{galanaki2018thunderstorm, Foote1979ResultsExperiment} which makes them hard to detect and predict.

The damages produced by intense winds, lightning and tornadoes have significant social impacts and incur significant liability for power grid operators. Overhead lines which are still widely used in rural areas are particularly prone to weather events. During the year 2017, 78 percent of all outages were caused by extreme weather events; 50 percent of these outages are caused by strong winds, 45 percent by ice and snow load, 5 percent by lightning and rest by other weather events \cite[p.~20]{Oy2017KESKEYTYSTILASTOi}. Extreme weather events cannot be prevented but power grid operators can minimise the effect of weather-induced damages. For example, they can up-level workforce in relevant areas when bad weather is anticipated.

Since weather-caused damages incur a significant economic loss, a lot of effort has been put into studying efficient prediction of impacts of extreme weather events. Blackouts caused by large scale hazards such as hurricanes have been studied in e.g. \cite{Guikema2014PredictingPlanning,Guikema2010PrestormSystems,Nateghi2014PowerModels,Han2009ImprovingModels,Wang2017ASystems,allen2014application,chen2016fuzzy, He2017NonparametricNetwork, Liu2018SceneNetwork}. In contrast, we focus on more localised phenomenons related to convective storms. The authors of \cite{li2015spatio} present an outage prediction method based on static areas and the authors of \cite{singhee2017probabilistic} ennoble this work to take power grid topology into account. \cite{shield2018predictive} uses Random Forest Classifier to a regular grid to create power outage prediction. In \cite{Zhou2006ModelingLines}, the authors compare a Poisson regression model and a Bayesian network model in the task of modelling failure rates in overhead distribution lines. These methods exploited data from ground weather stations and lightning detection network with a daily time interval. Kankanala et al. have experimented regression models \cite{kankanala2011regression} and multilayer perceptron (MLP) neural network \cite{Kankanala2012EstimationNetwork} along with ensemble learning \cite{kankanala2014adaboost} to predict outages caused by wind and lightning in overhead distribution systems. Their methods are based on data from nearby weather stations. Bayesian outage probability (BOP) model predicting power outages has been discussed in \cite{Yue2018AData}. They combined weather radar information from several sources to a geographically unified grid. Authors of \cite{cintineo2014empirical} propose a method to forecast a probability that developing thunderstorm will produce severe weather. The method consists of creating spatial objects from satellite and weather radar data, tracking them and classifying them to be hazardous or non-hazardous with Na\"ive Bayesian classifier. The method is focused on predicting tornadoes, severe wind gusts, and hailstones and is aimed to provide a tool to weather forecaster. The problem of storm cell identification and tracking has been studied thoroughly also in \cite{rossi2015object}. 

We propose a novel method to predict the impact of severe convective storms on the power grid. Our method combines storm cell identification and tracking developed in \cite{rossi2015object} with state of the art machine learning techniques. The storm cell identification and tracking used in this work are related to \cite{cintineo2014empirical}. However, \cite{cintineo2014empirical} considers a binary classification for weather forecasters whereas we predict the short-term damage potential specifically for a power grid. Similar to \cite{Zhou2006ModelingLines, kankanala2011regression}, we use data produced by weather stations and lightning detection network and combine them with parameters derived from weather radar data as used in \cite{Yue2018AData}. Moreover, since our method is based on identified storm cells, we are able to use also parameters characterising the storm cell itself. This provides much more accurate spatial and temporal resolution than weather stations and more information about the whole storm instead of any individual point. 

This paper is organised as follows: In Section \ref{sec:problem-formulation}, we formulate a problem as a classification problem. In Section \ref{sec:Proposed Method} we first discuss using an object-based approach in predicting power outages and propose two alternative classification methods of random forest classifier (RFC) and deep neural network classifiers to predict amount damage. Some illustrative numerical experiments based on historical data collected by the Finnish Meteorological Institute (FMI) are discussed in Section \ref{sec:Experiments} followed by results in Section \ref{sec:Results}.

\section{Problem Formulation} 
\label{sec:problem-formulation}

We model outage prediction as a supervised learning problem with occurred power outages as labels and weather conditions as features. Outage data and power grid description are fetched from two power grid operators. The data set contains in total of 33 858 outages. It is notable that actual damage may happen to any point in the power grid, but outages are always reported at nearby transformer nodes of the power grid. One physical damage may also turn down several transformers. Unbroken transformers can later be taken into use remotely by power grid operator (without repairing the actual damage). One physical damage can thus be reported as several outages in the power distribution network.

\begin{figure*}[ht]
\centering
\includegraphics[width=0.7\textwidth]{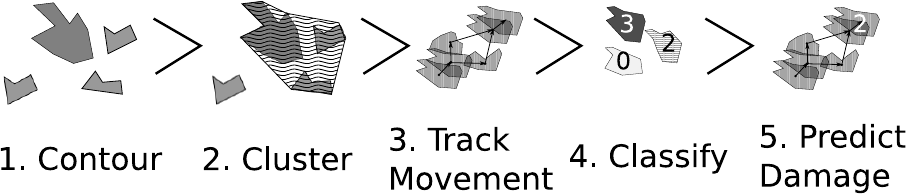}
\caption{Overall process: 1) find storm cells by contouring CAPPI images, 2) cluster storm cells 3) track storm cell movement 4) classify clusters based on their damage potential 5) predict future outages based on predicted location and class of the cluster. Steps 1-3 are first introduced in \cite{rossi2015object}.}
\label{fig:process}
\vspace{-10pt}
\end{figure*}

\begin{figure}[ht]
\centering
\includegraphics{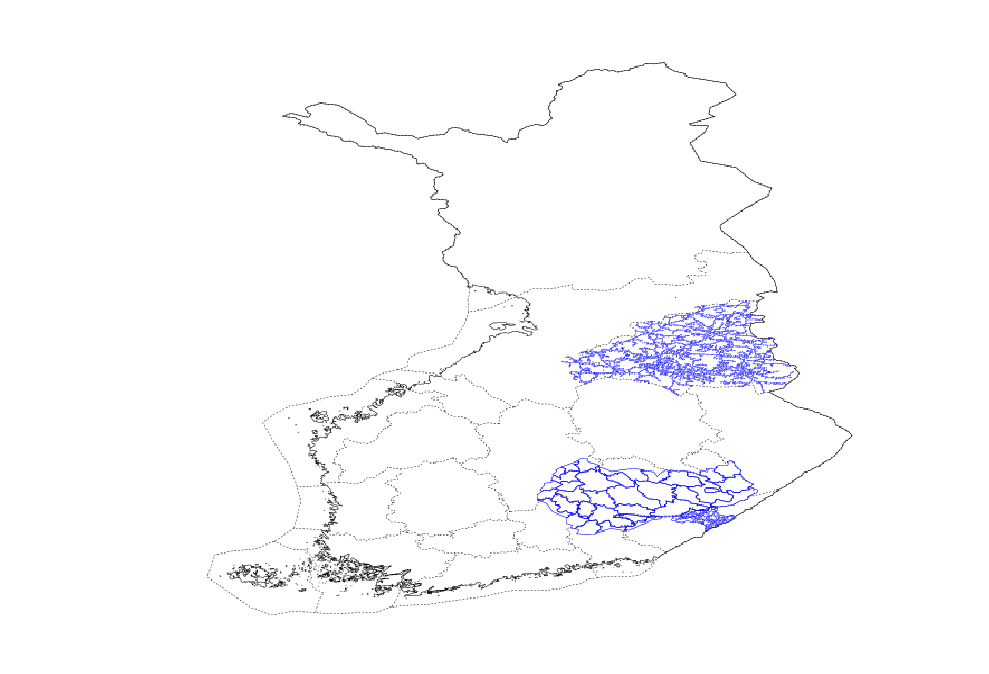} 
\caption{Spatial coverage of power grid information available in this project. Soft lines represents province districts and darker blue lines power grids.}
\label{fig:network_area}
\end{figure}

\begin{figure}[ht]
\centering
\subfloat[\label{fig:outage_heatmap_loiste}]{%
\includegraphics{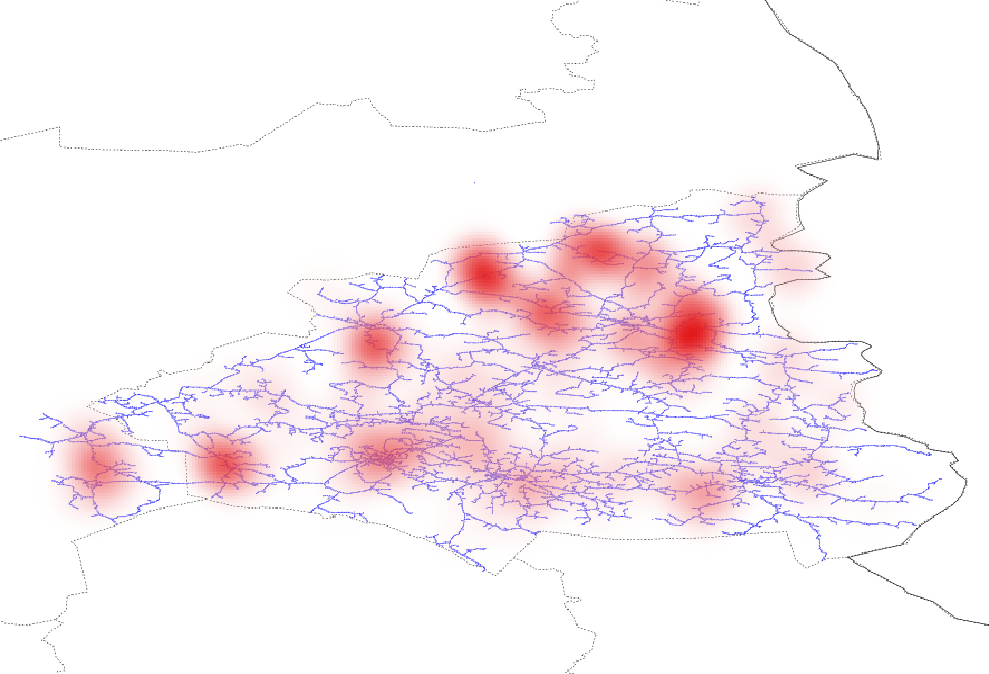}} \\
\subfloat[\label{fig:outage_heatmap_jse}]{%
\includegraphics{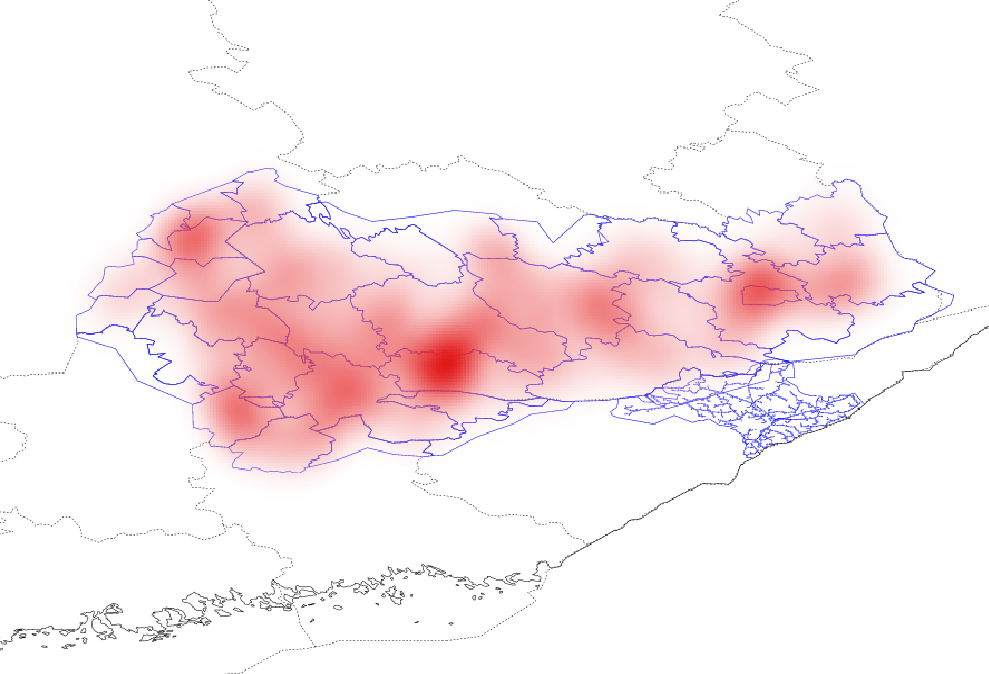}}
\caption{Spatial distribution of outage data. Darker red area represents more outages between 2012 and 2017. (a) Outage heatmap for Loiste network (Northern area). (b) Outage heatmap for JSE network (Southern area).}
\vspace{-10pt}
\end{figure}

\begin{figure}[ht]
\centering
\includegraphics{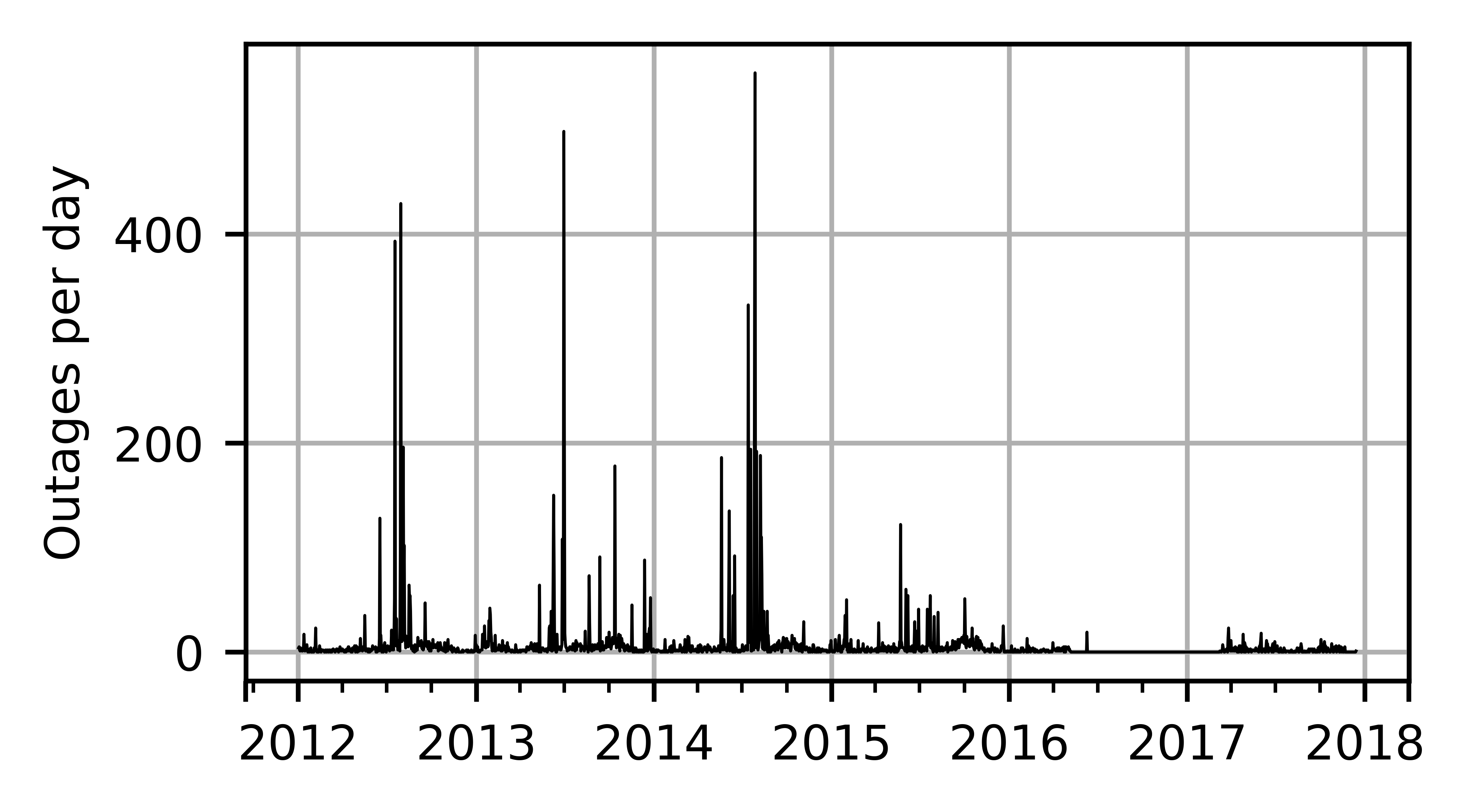} 
\caption{Amount of outages per day on the whole area.}
\label{fig:outage_timeseries}
\vspace{-10pt}
\end{figure}

Spatial coverage of the data is shown in Fig. \ref{fig:network_area}. Heatmaps of all outages recorded between 2012 and 2017 are shown in Fig. \ref{fig:outage_heatmap_loiste}. and \ref{fig:outage_heatmap_jse}. It is notable that outages are not distributed evenly over the area. Instead, both areas contain ``hotspots'' where outages are more common than elsewhere. Outages also occur very unevenly in time. Fig. \ref{fig:outage_timeseries} shows amounts of outages per day in the whole area. One can see that most outages have happened during only a few days. It is also notable that although the worst peaks in outages get a place during summers, there are minor peaks also during winters. These outages are most probably caused by wet snow load on trees and wires \cite{Makkonen2010SimulatingModel}. The method considered in this paper is unable to address these cases.

We cast the problem of recognising a damage potential of the storm to as categorisation problem and propose two alternative methods for the task. In particular, we categorise storms into four classes based on how much damage they are expected to cause for the power grid during one time step. A storm at any given time step is defined as a “sample” in the remainder of this paper. The storm cells are assigned to a class based on how large share of transformers under the storm is without electricity. That is, a number of transformers in the whole network do not effect on the classification. We use four classes, described in Table \ref{table:classes}. The particular choice of these classes aims to provide a simple 'at glance' view which is convenient for the end user (power grid operator). 

\setlength\e{\dimexpr .15\columnwidth -2\tabcolsep}
\setlength\ee{\dimexpr .50\columnwidth -2\tabcolsep}
\begin{table}[ht]
    \centering
    \caption{Class definitions of the storm cells.}
    \label{table:classes}
    \begin{tabular}[width=0.5\columnwidth]{ p{\e} p{\ee} } 
    \textbf{Class} & \textbf{Share of transformers } \\
    \hline
    0 & no damage \\ 
    \hline
    1 & 0 - 10 \% \\ 
    \hline
    2 & 10 - 50 \% \\ 
    \hline
    3 & 50 - 100 \% \\ 
    \end{tabular}
\end{table}

The data are very imbalanced as most of the storm cells are not powerful enough to cause harm to the power grid. We depict a histogram of the target classes contained in the training data set in Fig. \ref{fig:y_hist_syntetic}. In particular, we have 872 801 (98,5 \%) samples of class 0 (no harm), 5183 (0.6 \%) samples of class 1, 4538 (0.5 \%) samples belonged to class 2 and only 3499 (0.4 \%) belong to class 3 (most harmful).

\section{Proposed Methods}
\label{sec:Proposed Method}

We used weather data collected by FMI during the years 2012 to 2017. Data are collected from weather radars, weather stations, and lightning detection. The data are pre-processed as introduced in \cite{rossi2015object}. First, we identify storm cells by contouring weather radar reflectivity composite constant altitude plan position indicator (CAPPI) images with a solid 35 DBZ threshold. The particular chosen threshold enables detecting full storm systems like multicellular storms \cite{Rossi2013Real-timeData, Dixon1993TITAN:Methodology}. The radar images have 250 meters spatial resolution and 5 minutes update interval. Anomaly detection and removal with methods described in \cite{Peura2002ComputerRemoval} has been applied to the radar images as part of FMI operational image processing. Contoured storm cells are stored as geographical objects into a PostGIS database.

After contouring, we apply the GDBSCAN method \cite{sander1998density} (generalised form of DBSCAN \cite{ester1996density}) to cluster the contoured objects. Within the DBSCAN method, a storm cell is considered as a core point if the area sum of its nearby storm cells exceeds the given area threshold. The storm cells that are within the neighbourhood of a core point (inside given radius) but do not fulfil the minimum area criteria are considered as outliers. Together with their outliers, connected core points form a cluster. Storm cells that do not fulfil the area criteria of core points and are too far from any core point are regarded as noise. In this study, the area limit was set to $20~{\rm km}^2$ and the neighbourhood radius was set to $2~{\rm km}$. Different parameter values have previously been evaluated subjectively and used in \cite{Rossi2015KalmanStorms}.

After clustering, we track and nowcast movement of the storm cells using method originally introduced in \cite{Rossi2008AAnalysis}:  
first, we interpolate clusters of previous time step to the current time step using optical flow \cite{horn1981determining} with Lucas-Kanade method \cite{Bouguet2001PyramidalAlgorithm}. If interpolated and current time step clusters overlap over the required threshold, we consider them connected. Predicting movement of the storm cells is done by Kalman filtering based method introduced in \cite{Rossi2015KalmanStorms}. The prediction is done for a time horizon of 2 hours ahead with a time resolution of 5 minutes. Every storm cell is identified with a globally unique identifier so that the whole lifecycle of the storms can be tracked. In particular, all overlapping storm cells in the same cluster are assigned with the same identifier. The unique identifier also makes the storms easily referable afterwards. 

The pre-processing produces a training set where each sample is characterised by in total of 16 features listed in Table \ref{table:input_data} and a target class based on the number of occurred outages. The features contain parameters fetched directly from 2-dimensional CAPPI radar images like area, age, and radar reflectivity (DBZ) parameters. The storm center is also taken into account because the characters of forest vary significantly in different parts of Finland. Several ground observations fetched from weather stations under storm path are used as well. Because the location of outage reports contains a lot of inaccuracy and devastating phenomenons such as wind and lightning may occur also outside the cluster area, ground observation and outage search area are extended by buffering storm geometry with 0.1 degrees.

\setlength\e{\dimexpr .35\columnwidth -2\tabcolsep}
\setlength\ee{\dimexpr .65\columnwidth -2\tabcolsep}
\begin{table}[ht]
    \centering
    \caption{Used input features}
    \label{table:input_data}
    \begin{tabular}[width=\columnwidth]{ p{\e} p{\ee} } 
    \textbf{Feature} & \textbf{Explanation} \\
    \hline
    Area & Area covered by the storm cell \\ 
    \hline
    Age & Age of the storm \\ 
    \hline
    Lightning density & Lightning density under storm cell \\ 
    \hline
    Max DBZ & Maximum radar reflectivity of the storm cell (spatially). Represents maximum rain intensity. \\
    \hline
    Min DBZ & Minimum radar reflectivity of the storm cell (spatially). Represents minimum rain intensity. \\
    \hline
    Mean DBZ & Mean radar reflectivity of the storm cell (spatially) \\
    \hline
    Median DBZ & Median radar reflectivity of the storm cell (spatially)\\
    \hline
    Std of DBZ & Standard deviation of radar reflectivity of the storm cell (spatially)\\
    \hline
    Lat & Storm center latitude \\
    \hline
    Lon & Storm center longitude \\
    \hline
    Temperature & Air temperature from ground\\ observations \\
    \hline
    Pressure &  Air pressure from ground\\ observations \\
    \hline
    Wind speed &  Wind speed from ground\\ observations \\
    \hline
    Wind direction &  Wind direction from ground\\ observations \\
    \hline
    Precipitation amount &  Precipitation amount from ground observations \\
    \hline
    Snow depth &  Snow depth from ground\\ observations \\
    \end{tabular}
\end{table}

Our overall process is described in Fig. \ref{fig:process}. A novel approach of this work is to combine classification methods to identified and tracked storm cell clusters to create power outage prediction. Instead of predicting (predefined limits of) weather variables, we focus on the impact of weather and let the classification method to deduce relevant variables and limits. Applying classification to (clustered) storm cells instead of using predefined areas or grid-based methods provides more accurate spatial and temporal resolution. Moreover, we argue that handling storms as geographical objects allows us to better characterise damage potential. Implementing classification to clusters instead of individual storm cells enables capturing the whole life cycle of the storm system \cite{Rossi2014AnalysisFinland}. We also argue that classification instead of regression better captures areas of interest from power grid operator point of view. 

We created two alternative methods for classification: Random Forest Classifier (RFC) \cite{breiman2001random} and multilayer perceptron (MLP) neural network \cite{Goodfellow-et-al-2016-mlp}. Random forest classifier is an ensemble method which forms a decision tree based on randomly selected samples from train data. The method is reported to work well for the imbalanced data \cite{Zhang2011AnData, Pal2005RandomClassification} and is hence very interesting candidate for this particular application. For the training of the RFC, we used the Gini impurity as loss function, i.e.
\begin{equation}
G = -\sum_{i=1}^{n_c}(p_i(1-p_i))
\vspace*{-1mm}
\end{equation}
where $n_c$ is the number of classes and $p_i$ is the share of $i^{th}$ class in the tree.

\begin{figure}[ht]
\centering
\includegraphics[width=\linewidth]{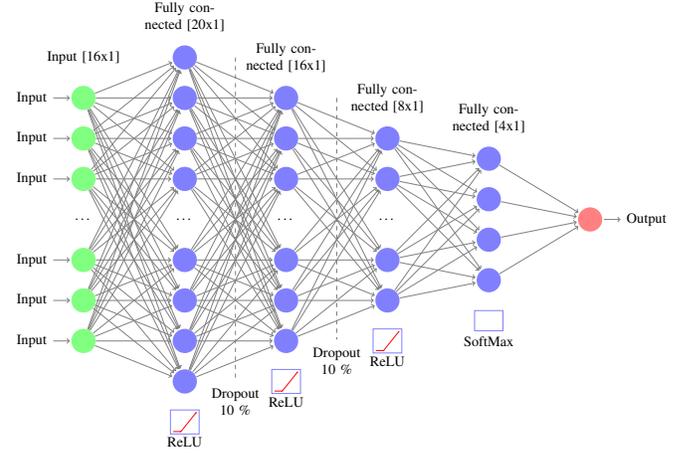} 
\caption{Network structure for classification task}
\label{fig:model_class}
\end{figure}

As an alternative to RFC, we also implemented a classifier based on a multilayer perceptron (MLP) neural network \cite{Goodfellow-et-al-2016-mlp}. The network structure, among hyperparameters, was searched by trial and error. The best solution we were able to find is described in Fig. \ref{fig:model_class}. The first layer contains 20 nodes wide dense layer. In the following layers, the number of nodes is reduced to 16, 8, 4 and finally to 1 node. The first three dense layers use the rectified linear unit (Relu) as activation and dropout regularisation layers are included after first and second dense layers. In the final layer, we used the ``Softmax'' activation function in order to obtain the predicted class probabilities in the output layer. For MLP, we used the cross-entropy loss function \cite{Goodfellow-et-al-2016-dropout}. This loss is defined as 
\begin{equation}
H(p,q) = -\sum_{x}(p(x)\log{(q(x))})
\vspace*{-1mm}
\end{equation}
where $p(x)$ is a probability distribution of true labels and $q(x)$ is a probability distribution of predicted labels. Categorical entropy is a good default choice and it has an optional advantage that different classes can be easily preferred by giving different weights for the classes.

\begin{figure}[t]
\centering
\includegraphics[width=0.9\linewidth]{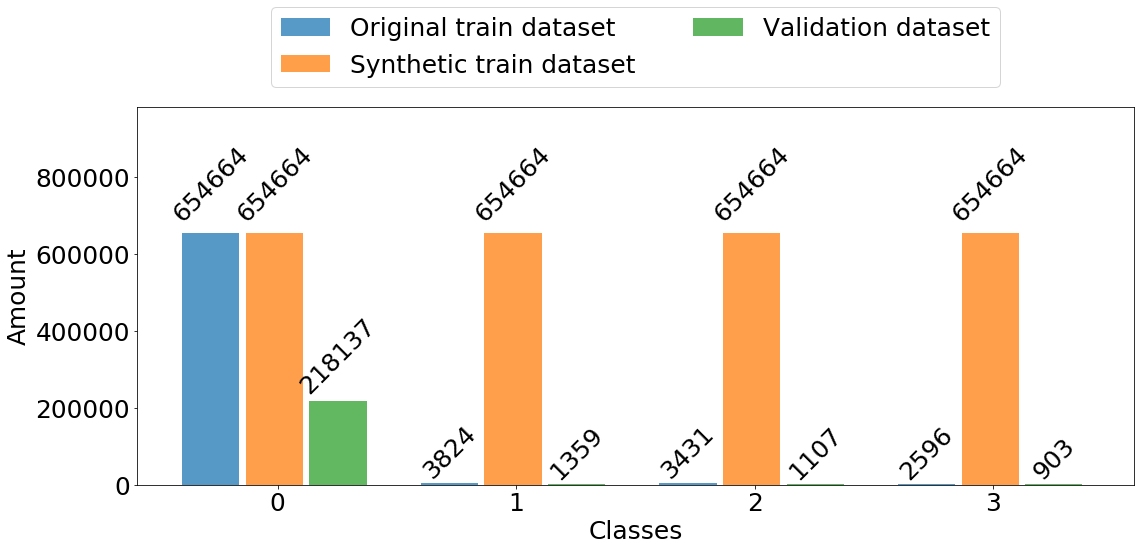} 
\caption{Histogram of classes in original data set divided to train and validation set (75 \% / 25 \% ratio) and synthetic data set after generating new samples by SMOTE.}
\label{fig:y_hist_syntetic}
\vspace{-10pt}
\end{figure}

We combined the classifiers with the synthetic minority over-sampling technique (SMOTE) \cite{chawla2002smote} to handle imbalanced data. The method generates new training samples in the vicinity of the original training samples by interpolating their $k=5$ nearest neighbours (in the feature space) as following: 
\begin{equation}
x_{new} = x_i + \lambda \times(x_{zi} - x_i)
\vspace*{-1mm}
\end{equation}where $x_i$ is an original minority class sample, $x_{zi}$ is one of $x_i$'s $k$ nearest neighbour and $\lambda$ is random variable drawn uniformly from the interval $[0,1]$. The synthetic data set generated by SMOTE contains data points with a balanced distribution of classes (see Fig. \ref{fig:y_hist_syntetic}).

\section{Numerical Experiments}
\label{sec:Experiments}

We divided the data set into training and validation set with share of 75 \% and 25 \% respectively. SMOTE over-sampling was performed only for the training set. For RFC, optimal hyperparameters were obtained with random search cross-validation \cite{bergstra2012random}. We used F1 macro average score for cross-validation so that all classes are valued equally in imbalanced data set. To be more specific:

\begin{equation}
\label{eq:F1}
F1_{macro} = \frac{1}{N}\sum_{\lambda=1}^N\big(\frac{precision_\lambda \times recall_\lambda}{precision_\lambda + recall_\lambda}\big)
\vspace*{-1mm}
\end{equation}
\begin{equation}
\label{eq:precision}
precision_\lambda = \frac{t_{p\lambda}}{t_{p\lambda}+f_{p\lambda}}
\end{equation}
\begin{equation}
\label{eq:recall}
recall_\lambda = \frac{t_{p\lambda}}{t_{p\lambda}+f_{n\lambda}}
\end{equation}
 
where $t_{p\lambda}$ is amount of true positive samples, $f_{p\lambda}$ false positives and $f_{n\lambda}$ false negatives in class $\lambda$. F1 scores of three best evaluation varied relatively much (0.70, 0.68 and 0.66) which indicates that the model is sensitive to the hyperparameters. The best obtained parameter set is listed in table \ref{table:rfc_hyperparameters}. The classification accuracy obtained for the training set was around 98 \% and for the validation set up to 88 \%. Thus, 
the RFC tends to slightly over-fit the training data. 

\setlength\e{\dimexpr .20\columnwidth -2\tabcolsep}
\setlength\ee{\dimexpr .55\columnwidth -2\tabcolsep}
\begin{table}[ht]
    \centering
    \caption{Hyperparameters of the RFC classifier}
    \label{table:rfc_hyperparameters}
    \begin{tabular}[width=\columnwidth]{ p{\ee} p{\e} } 
    \textbf{Parameter} & \textbf{Value } \\
    \hline
    Number of trees in the forest & 200 \\ 
    \hline
    Max depth & unlimited \\ 
    \hline
    Minimum nro. of samples to split & 2 \\ 
    \hline
    Minimun nro of samples to leaf & 1 \\
    \hline
    Features to consider for split & 4 \\
    \hline
    Max nro of leaf nodes & unlimited \\
    \end{tabular}
\end{table}

\begin{figure}[ht]
\centering
\includegraphics{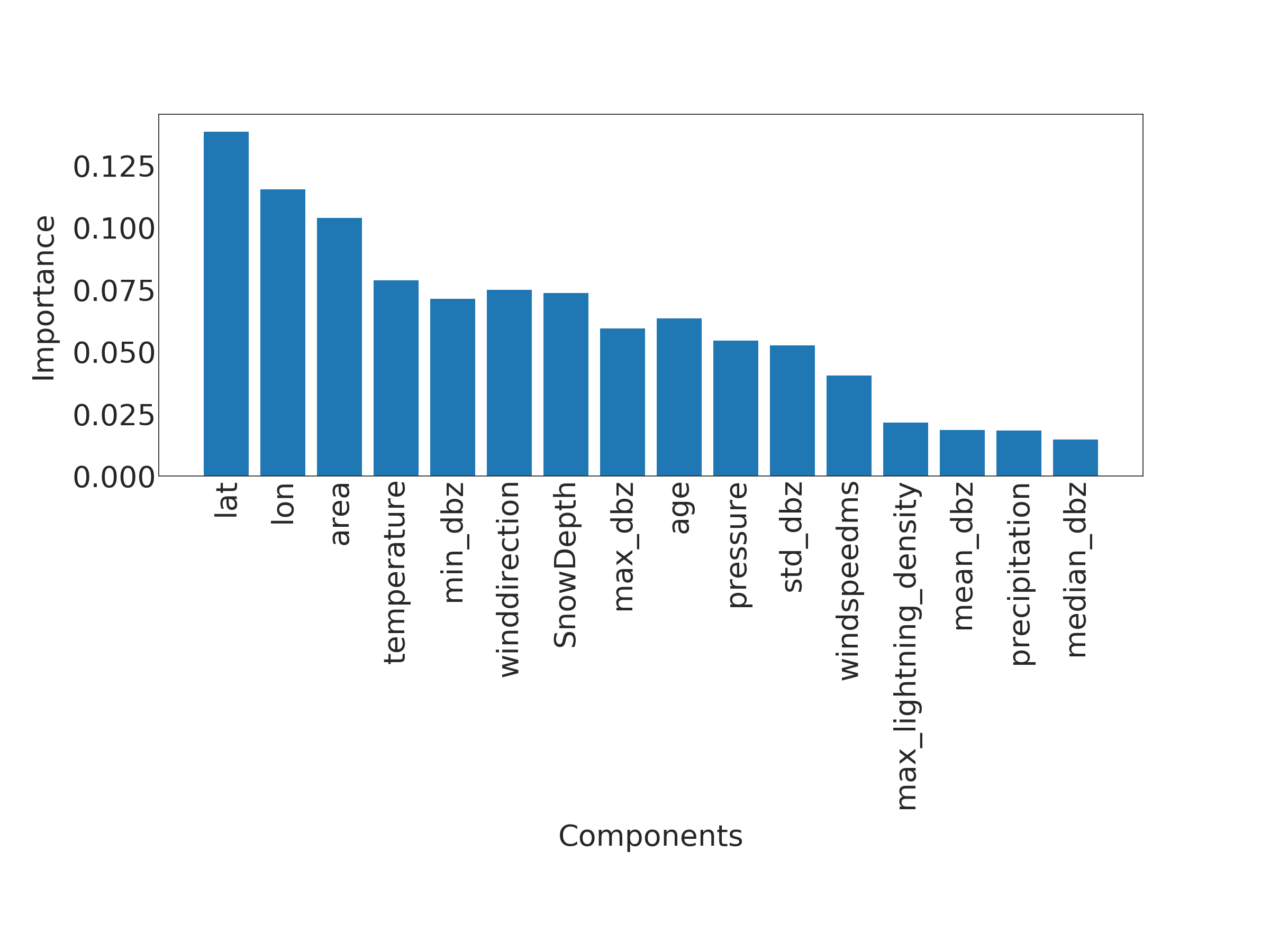} 
\caption{Feature importance in RFC model. The importance is defined as `gini importance'.}
\label{fig:rfc_feature_importance}
\end{figure}

One nice feature of RFC is that it provides relatively easy means to extract the importance of used features. The feature importance is plotted in Fig. \ref{fig:rfc_feature_importance}. The importance analysis indicates that 12 out of 16 used features have significance in the classification. One can see that the storm center (latitude and longitude) are by far the most important features which makes sense as occurred outages are condensed at certain areas. It is also notable that the most important feature fetched from weather radar is an area of the storm, not the intensity of precipitation (DBZ). Several ground observations such as temperature, wind, pressure, and snow depth have relatively high importance as well. We tried to fit the method with a different number of features. While differences were small, the best results were nevertheless obtained with all 16 features. 

\begin{figure}[ht]
\centering
\includegraphics[width=0.9\linewidth]{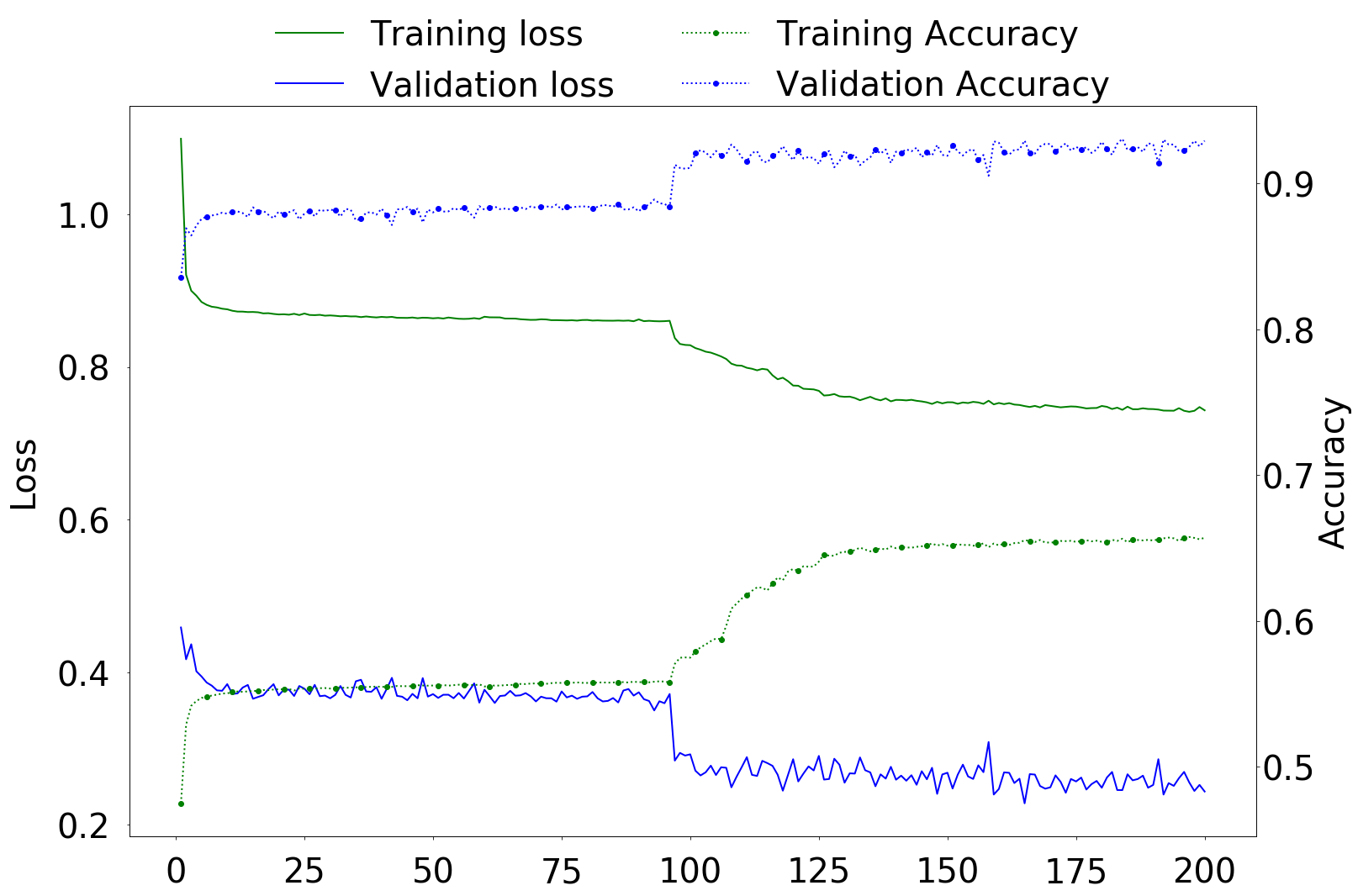} 
\caption{Training and validation metrics while training the MLP network. Validation performance is significantly better than training performance since training data set contains synthetically generated samples which do not exist in the validation set.}
\label{fig:nn_perf}
\vspace{-10pt}
\end{figure}

Finding an optimal setup for the MLP network was a significantly more challenging task. Both network topology and hyperparameters were searched by trial and error. We started from one hidden layer and added more layers as long as they improved the results. For each round, we evaluated several different sizes for hidden layers with different activation functions. Training and validation loss and accuracy of the final training are plotted in Fig. \ref{fig:nn_perf}. There is no sign of over-fitting and thus a quite low dropout probability was used for both dropout layers \cite{Goodfellow-et-al-2016-dropout}. The batch size was a compromise to provide enough performance in sufficient time. We used the Adam optimiser \cite{kingma2014adam} for training the model to avoid challenging points in the optimisation space. Final setup for hyperparameters is listed in table \ref{table:hyperparameters}.

\setlength\e{\dimexpr .20\columnwidth -2\tabcolsep}
\setlength\ee{\dimexpr .55\columnwidth -2\tabcolsep}
\begin{table}[ht]
    \centering
    \caption{Hyperparameters of the MLP classifier}
    \label{table:hyperparameters}
    \begin{tabular}[width=\columnwidth]{ p{\ee} p{\e} } 
    \textbf{Parameter} & \textbf{Value } \\
    \hline
    Batch size & 256 \\ 
    \hline
    Epoch count & 1000 \\ 
    \hline
    Dropout probability & 10 \% \\ 
    \hline
    $\alpha$ (learning rate) & 0.001 \\
    \hline
    $\beta_1$ (exp decay for momentum) & 0.9 \\
    \hline
    $\beta_2$ (exp decay for momentum) & 0.999 \\
    \hline
    $\epsilon$ (stability constant) & $10^{-8}$ \\
    \hline
    Initial decay & no decay \\
    \end{tabular}
\end{table}

Data contained a large number of incomplete samples as capabilities of weather stations varies a lot. Absent parameters were initialised to zero to ensure technical coherence of the data. The intuitive assumption would say that filtering those samples would be beneficial to gain better results. We created a new data set $\mathbf{X_{filt}}$ from samples which contained all parameters and used that to train the classification methods. The new data set contained 563 571 samples (63 \% of the original data set). Optimal hyperparameters for RFC with the new data set were re-optimised with random search cross-validation. For MLP, we used the same setup as for the full data set.

\subsection{Results}\label{sec:Results}

\setlength\e{\dimexpr .15\columnwidth -2\tabcolsep}
\setlength\ee{\dimexpr .38\columnwidth -2\tabcolsep}
\begin{table}[ht]
    \centering
    \caption{Metrics for different methods evaluated with validation set. 'MLP' and 'RFC' means corresponding methods while $\mathbf{X_{filt}}$' stands for samples without any missing values. Values with bold font are the best achieved values with each metric.}
    \label{table:results}
    \begin{tabular}[width=\columnwidth]{ p{\ee} p{\e} p{\e} p{\e} p{\e} } 
    \textbf{Metrics} & \textbf{MLP} & \textbf{MLP $\mathbf{X_{filt}}$} & \textbf{RFC} & \textbf{RFC $\mathbf{X_{filt}}$} \\
    \hline
    Accuracy & 93 \% & 85 \% & \textbf{100 \%} & 98 \% \\
    \hline
    AUC & 0.82 & 0.81 & 0.86 & \textbf{0.88} \\
    \hline
    Precision \\ micro average & 93 \% & 85 \% & \textbf{100 \%} & 99 \% \\
    \hline
    Precision \\ macro average & 29 \% & 31 \% & 66 \% & \textbf{67 \%} \\
    \hline
    Recall \\ micro average & 93 \% & 85 \% & \textbf{100 \%} & 99 \% \\
    \hline
    Recall \\ macro average & 65 \% & 67 \% & 75 \% & \textbf{79 \%} \\
    \hline
    F1 score \\ micro average & 93 \% & 85 \% & \textbf{100 \%} & 99 \% \\
    \hline
    F1 score \\ macro average & 32 \% & 34 \% & 70 \% & \textbf{72 \%} \\
    \end{tabular}
    \vspace{-10pt}
\end{table}

\begin{figure}[ht!]
\centering
\subfloat[\label{fig:confusion_matrix_rfc_full}]{%
\includegraphics{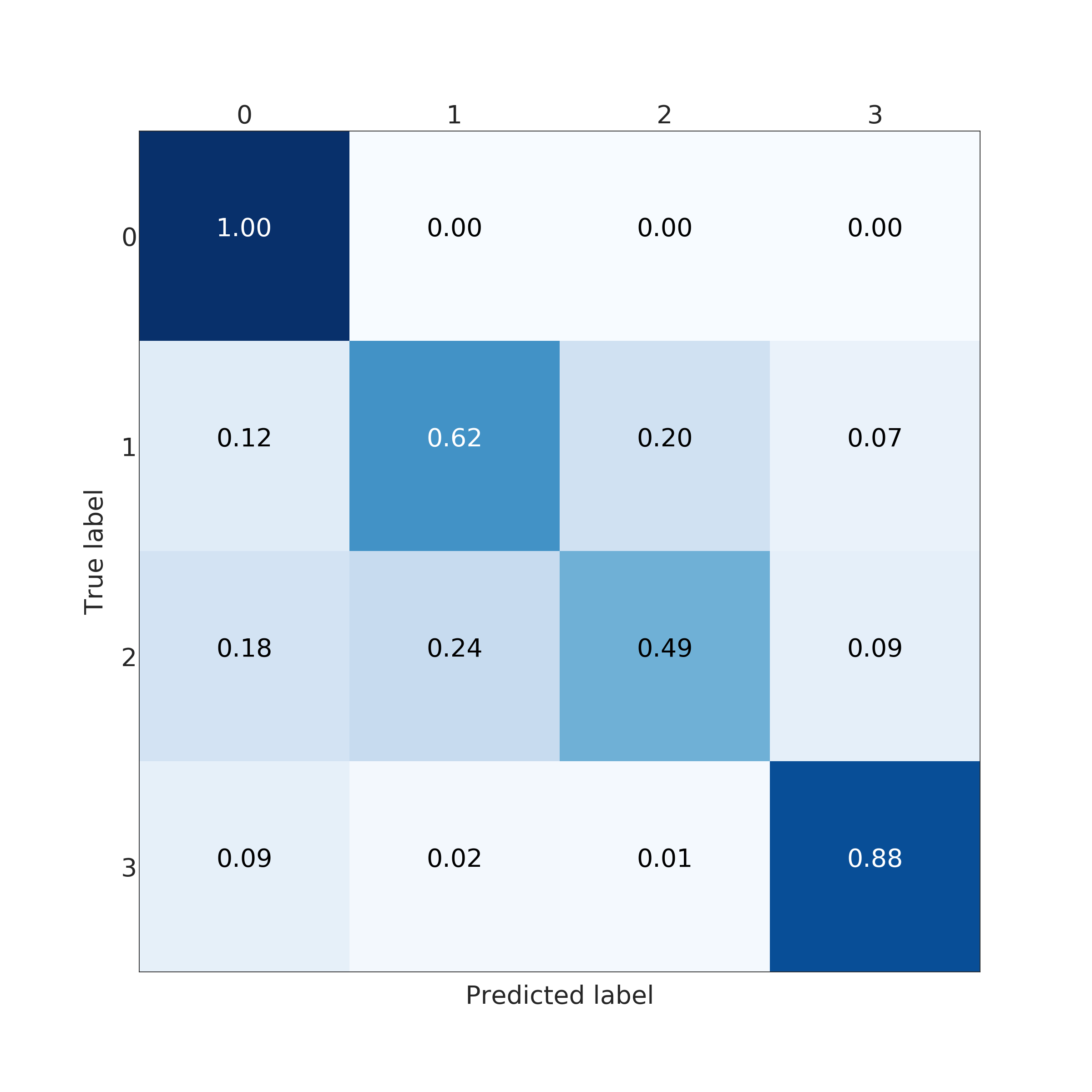}}\\
\subfloat[\label{fig:confusion_matrix_rfc_filtered}]{%
\includegraphics{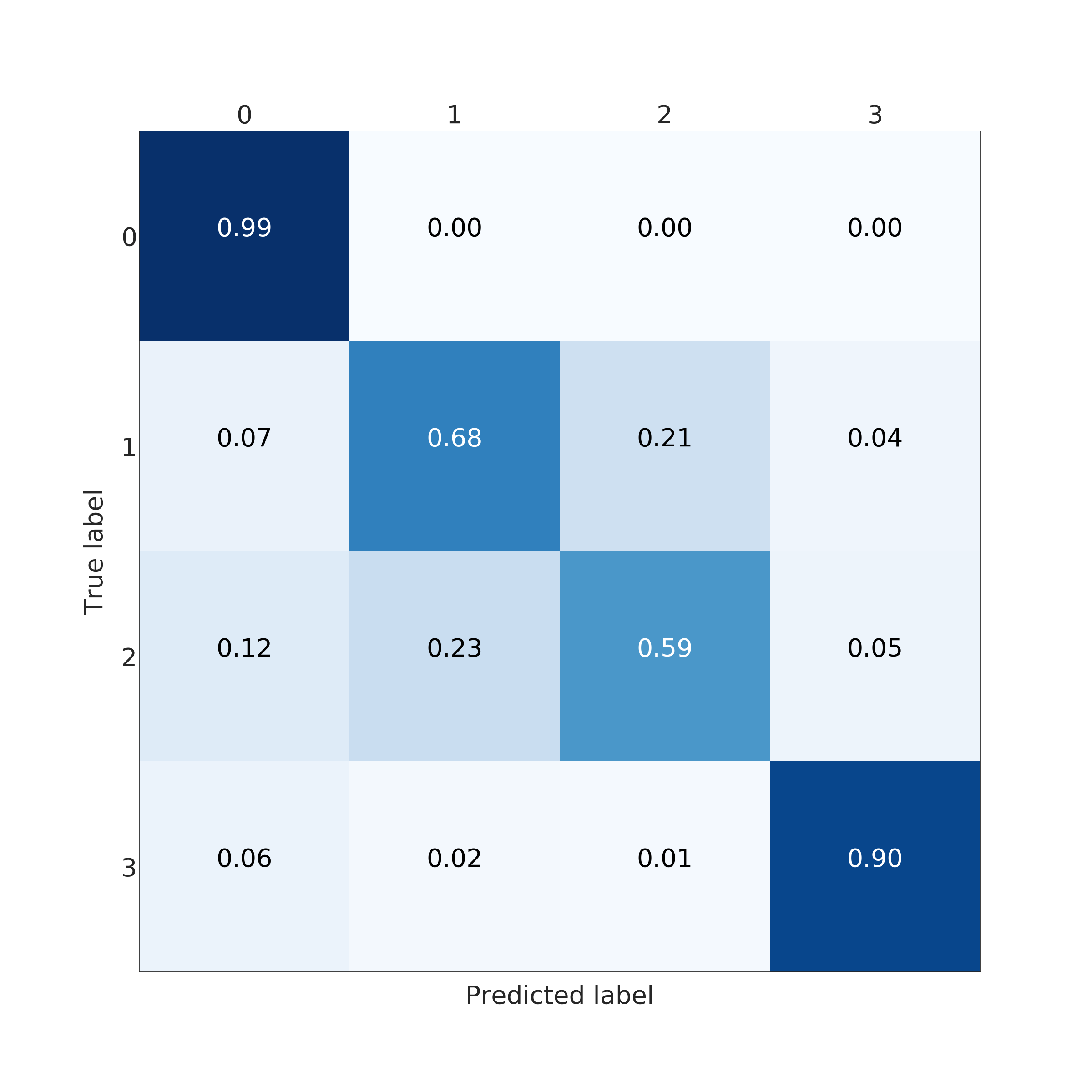}}
\caption{Normalised confusion matrices. (a) RFC with the full data set (b) RFC with the filtered data set. Each cell in the matrix represents the probability of predicted and true label combination. For example second cell in the first row gives a probability that used RFC predicts class 1 when true class is 0.}
\label{fig:confusion_matrix_rfc}
\end{figure}

\begin{figure}[ht!]
\centering
\subfloat[\label{fig:confusion_matrix_nn_full}]{%
\includegraphics{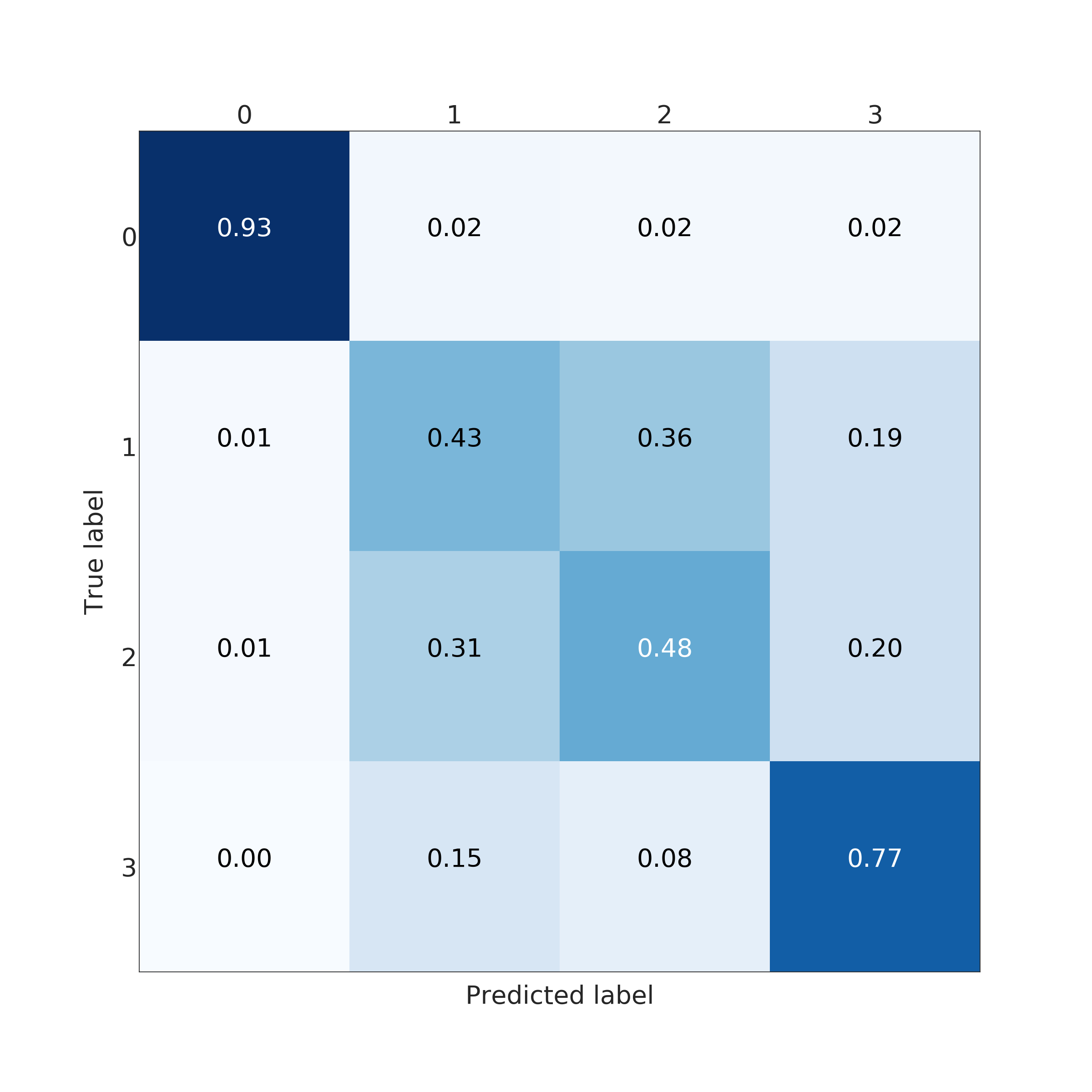}} \\
\subfloat[\label{fig:confusion_matrix_nn_filtered}]{%
\includegraphics{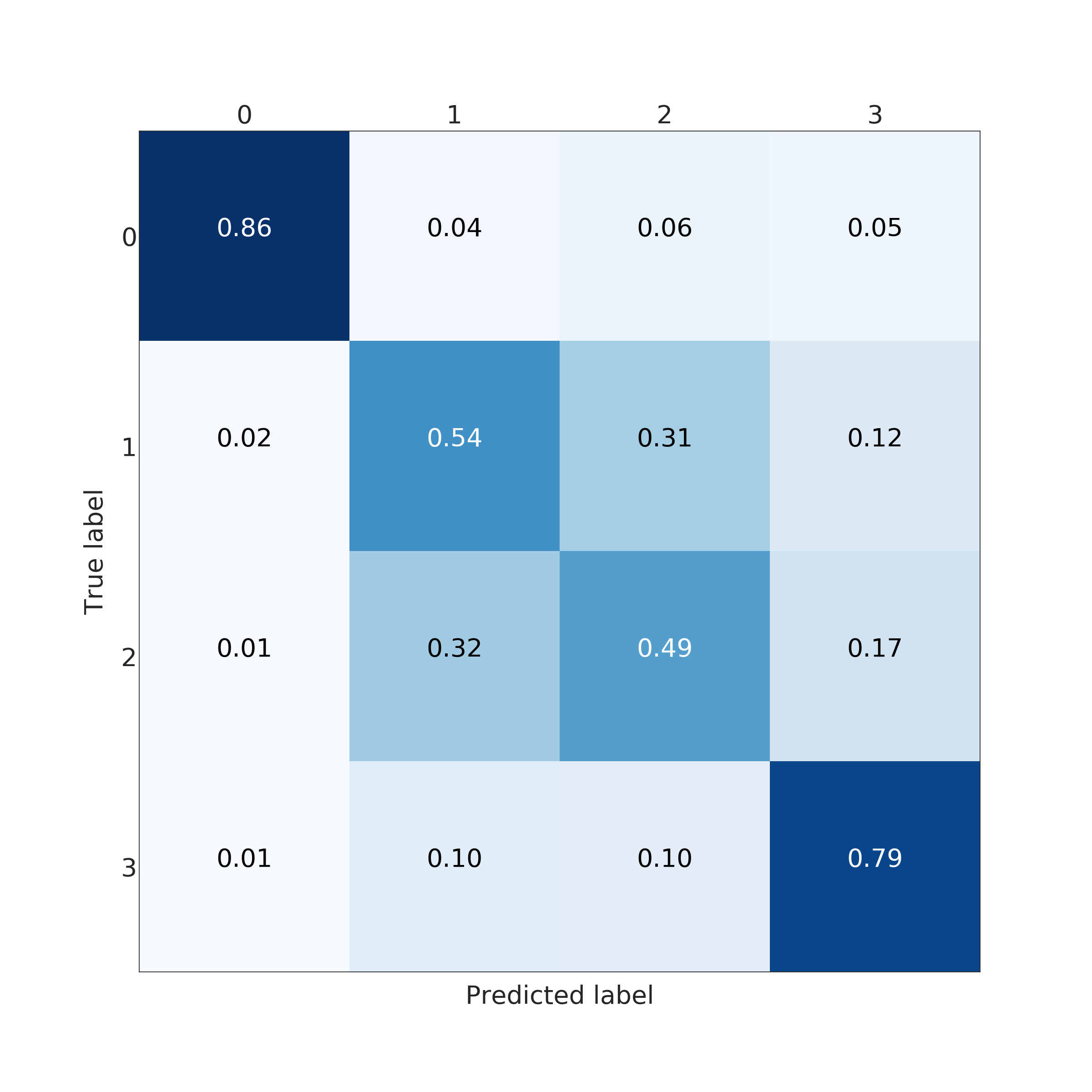}}
\caption{Normalised confusion matrices. (a) MLP with the full data set (b) MLP with the filtered data set. For example second cell in the first row gives a probability that used MLP predicts class 1 when true class is 0.}
\label{fig:confusion_matrix_nn}
\end{figure}

\begin{figure}[ht!]
\centering
\subfloat[\label{fig:prec-rec_rfc_full}]{%
\includegraphics{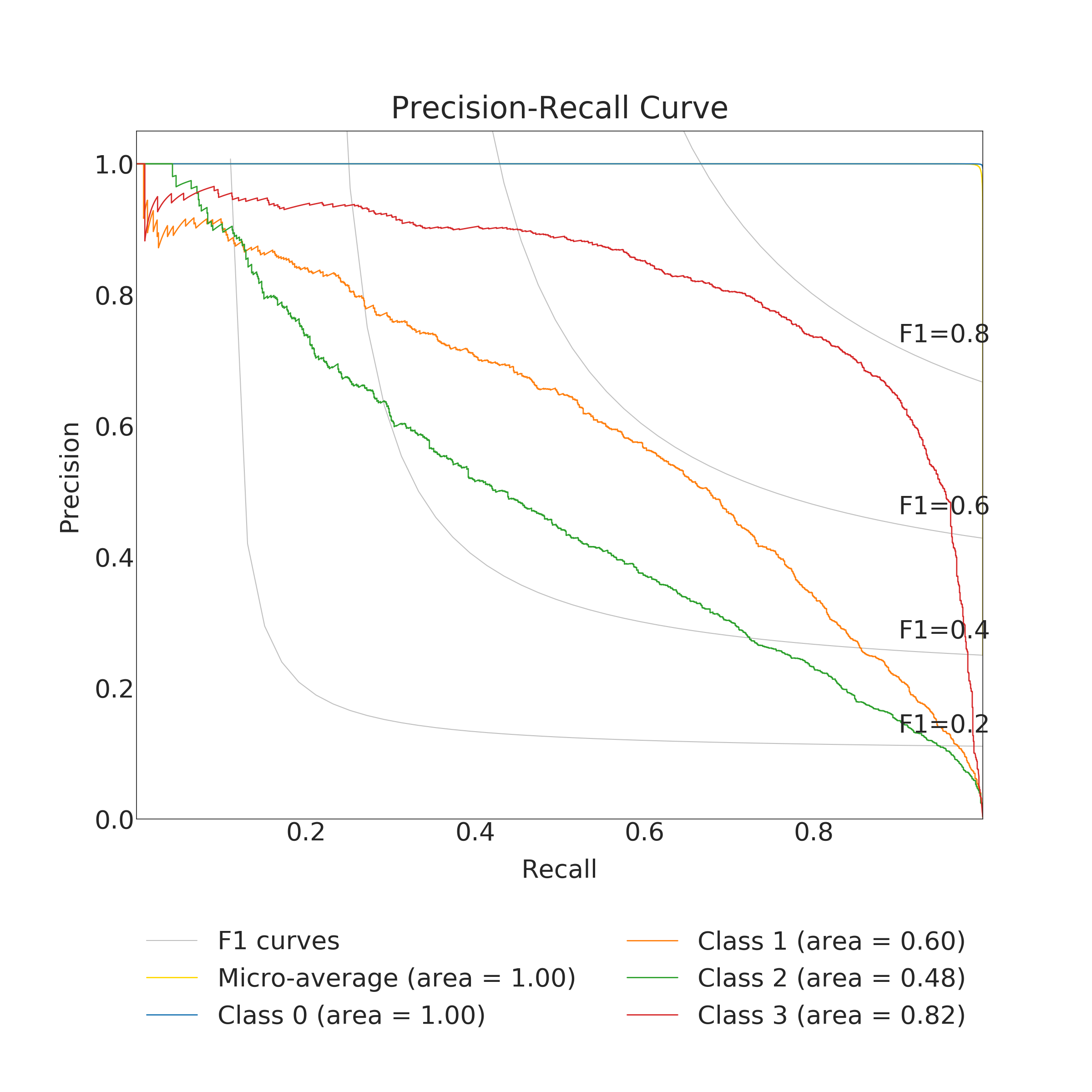}} \\
\subfloat[\label{fig:prec-rec_rfc_filtered}]{%
\includegraphics{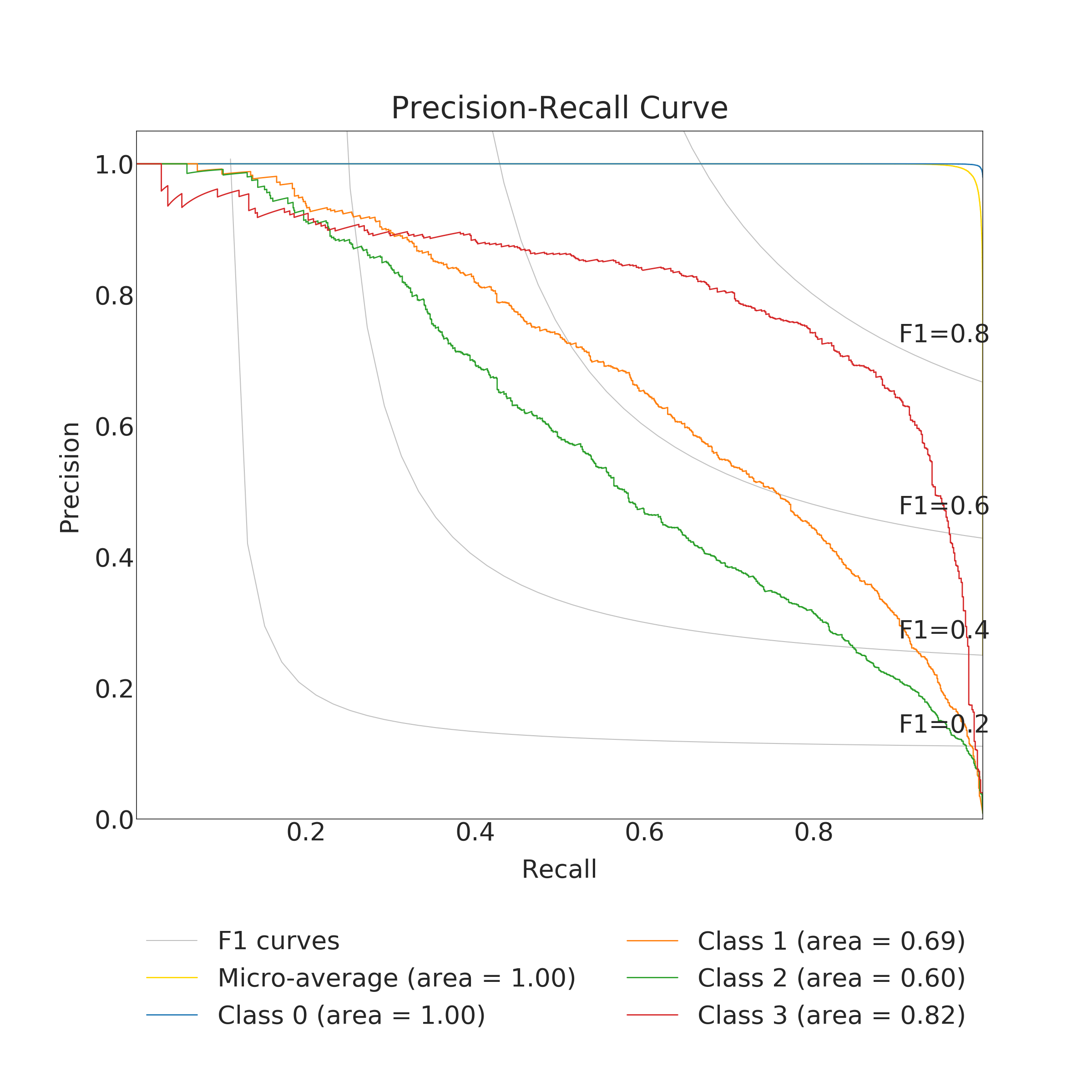}} 
\caption{Precision-Recall curves. (a) RFC with the full data set (b) RFC with the filtered data set}
\label{fig:prec-rec_rfc}
\end{figure}

\begin{figure}[ht!]
\centering
\subfloat[\label{fig:prec-rec_nn_full}]{%
\includegraphics{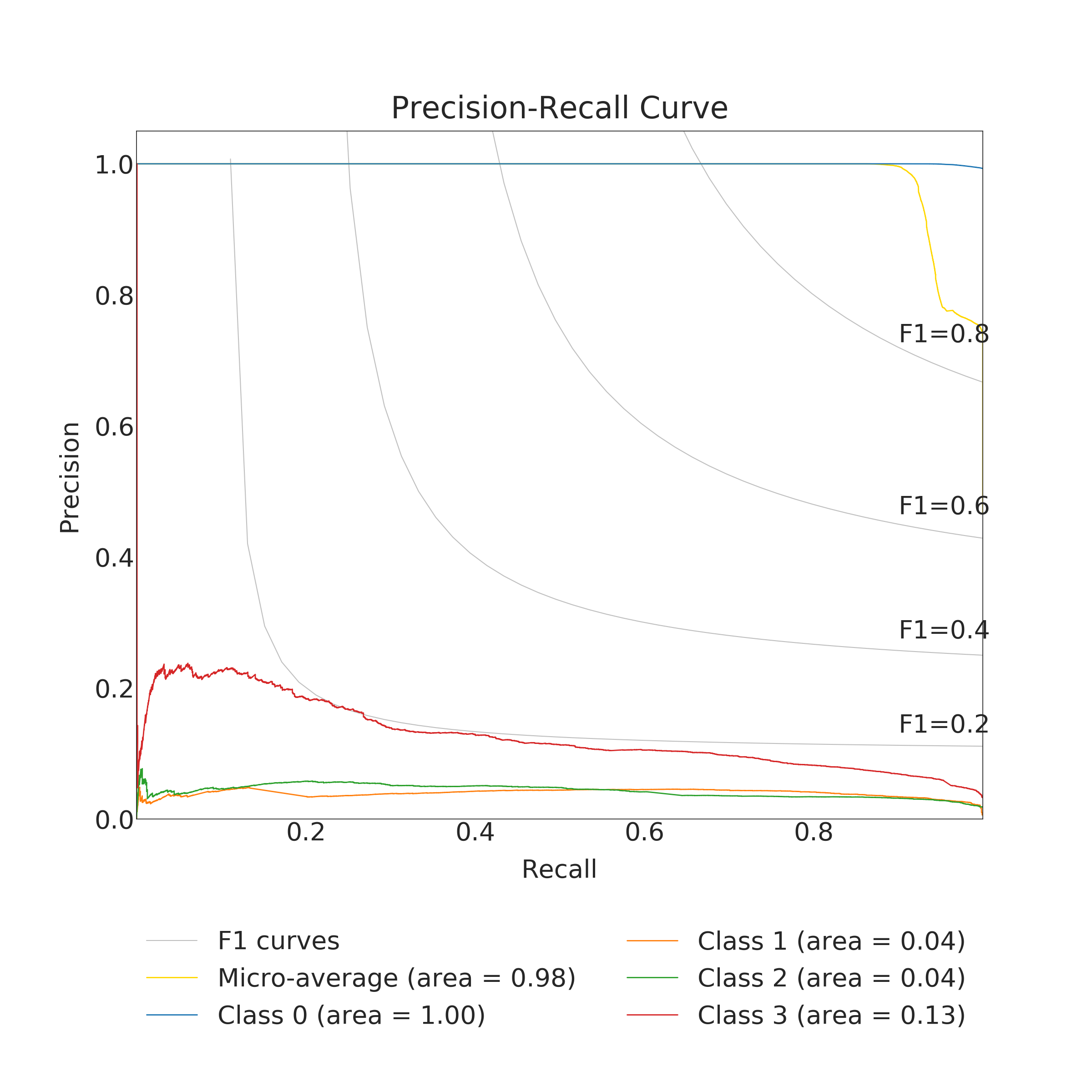}} \\
\subfloat[\label{fig:prec-rec_nn_filtered}]{%
\includegraphics{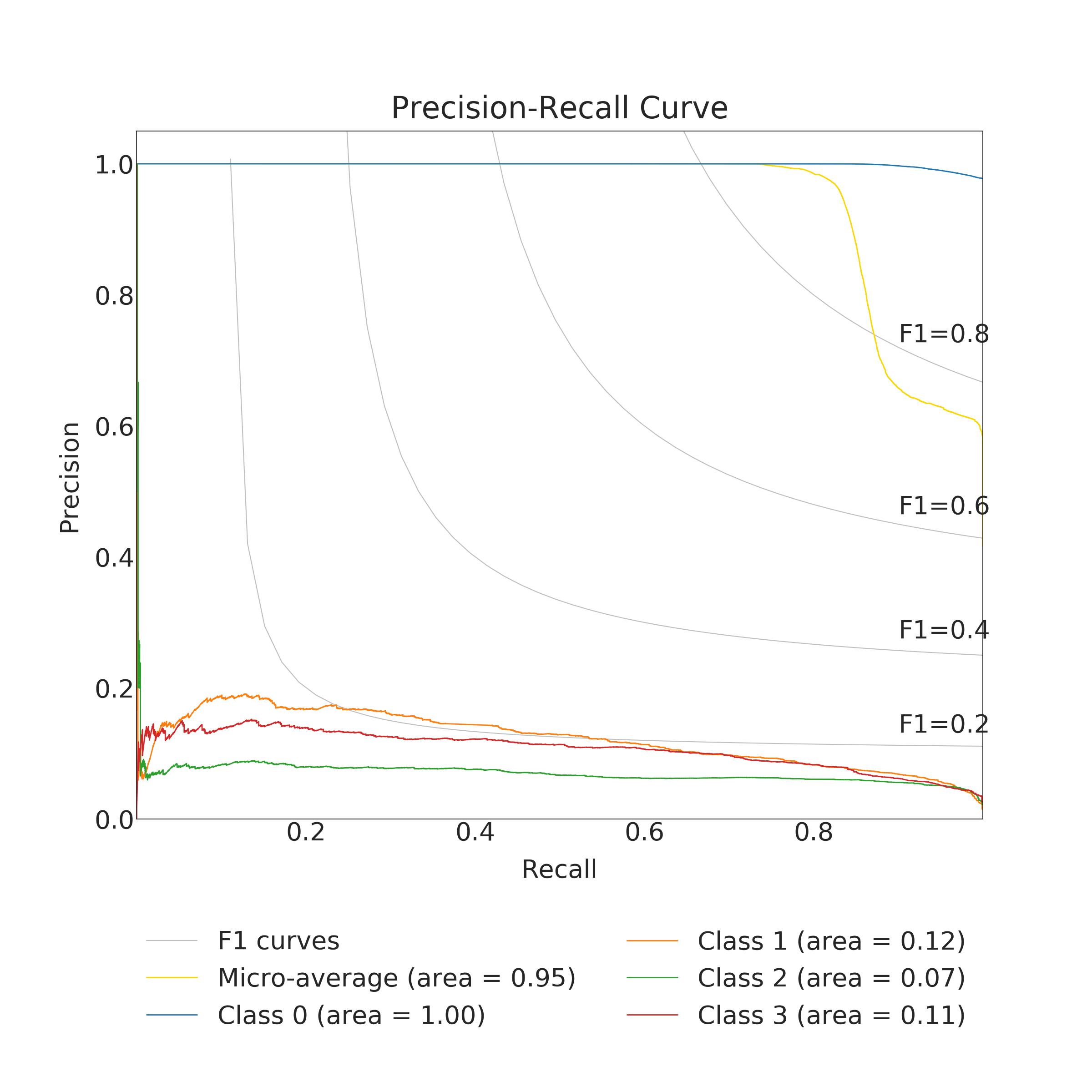}}
\caption{Precision-Recall curves. (a) MLP with the full data set (b) MLP with the filtered data set}
\label{fig:prec-rec_nn}
\end{figure} 

The results are shown in table \ref{table:results}. We evaluated performance with accuracy, Area Under the Curve (AUC) precision, recall and F1 score using both micro and macro average where applicable. The micro average is calculated from all samples from all classes. The macro average is calculated by taking an average of scores counted from each class independently. \cite[p.~679]{tsoumakas2009mining}. To be more specific, having binary metrics $M(t_p, f_p, t_n, f_n)$ where $t_p, f_p, t_n$, and $f_n$ are amount of true positive, false positive, true negative and false negative samples respectively, one can define macro average as:
\begin{equation}
    M_{macro} = \frac{1}{N}\sum_{\lambda=1}^N M(t_p, f_p, t_n, f_n)
\end{equation}
and micro average as:
\begin{equation}
    M_{micro} = M(\sum_{\lambda=1}^N t_p, \sum_{\lambda=1}^N f_p, \sum_{\lambda=1}^N t_n, \sum_{\lambda=1}^N f_n)
\end{equation} 
where $N$ is amount of samples. Thus micro average is more suitable metrics for imbalanced data.

In the end, RFC performed better for predicting the amount of damage. Its accuracy and a micro average of precision, recall and F1 score were nearly 100 \% (depending on used data set for training) but corresponding macro averages are significantly lower, varying from 66 to 79 \%. Differences between micro and a macro average of used metrics are explained by dominating class 0 (no damage) in the data set. Performance for individual classes can be best seen in confusion matrices (Fig. \ref{fig:confusion_matrix_rfc}. and \ref{fig:confusion_matrix_nn}). RFC is able to predict class 0 (no damage) with over 99 \% accuracy and class 3 (most harmful) with 90 \% accuracy but has troubles to distinguish classes 1 and 2 (68 to 59 \% accuracy respectively). Precision-recall curves are plotted in Fig. \ref{fig:prec-rec_rfc}. and \ref{fig:prec-rec_nn}. All models have extremely good precision (defined in equation \ref{eq:precision}) and recall (defined in equation \ref{eq:recall}) for class 0. Both RFC versions perform relatively well for class 3 but do not show proper skill for classes 1 and 2 having precision and recall mostly below 0.5. MLP works remarkably worse in terms of precision. 

Impacts of filtering samples with missing values are two-pronged. Using clean data ($\mathbf{X_{filt}}$) for training improved the AUC and the macro average of precision, recall and F1 score but decreased accuracy a little bit. RFC trained with clean data set performed a little bit better with classes 1, 2 and 3 but worse with dominating class 0. That is to say, using clean data in training yielded to a little bit more false alarms but less missed cases. Nevertheless, the differences were marginal. 

The results of the MLP classifier was significantly worse. While the MLP trained with full data set provided 93 \% accuracy for class 0 and 77 \% accuracy for class 3, accuracy for classes 1 and 2 are below 50 \%. The MLP provided especially poor precision and F1 score giving a high false positive rate compared to true positive rate. Using the full data set for training yielded to better results than the clean data set with the MLP.

Classification is fast with both classifiers. For example, 221 506 samples can be classified in about 5 seconds with two 3,3 GHz i5 CPUs. While optical flow is a lightweight algorithm, DBSCAN method has a $\mathcal{O}(n\log{}n)$ computational cost \cite{ester1996density}, which may be challenging during ``active days''. Performing the clustering in under five minutes before the next radar image is produced requires significant computing power. Our experience is that at least 16 core server is required for processing area covering Finland.

\section{Conclusions}
\label{sec:Conclusion}

This paper studied the application of RFC and MLP classifiers to the problem of predicting power grid outages caused by hazardous storm cells. The classification method was based on characteristics of the storm cell extracted from CAPPI weather radar images, related ground weather observations, and lightning detection information. 

Some illustrative numerical experiments based on weather data collected by FMI indicated that RFC can outperform deep MLP in predicting the amount of damage caused by storm cells. While MLP provided only poor performance, RFC showed very promising potential for the prediction task. Specially non-harmful and the most harmful cases were predicted with excellent accuracy.

This work suggests several interesting avenues for future research. Although used features already covered a quite wide range of environmental measurements, several promising but unutilised data sources still exist. Echo top information and speed of the storm could possibly give a good indication of its damage potential. Storm center seems to help the current classification method a lot but high importance of location prevents the method to be used in other areas without re-training it. The more generalised approach would maybe be to use the height of the forest as an input. 

One promising direction is to use more advanced models and methods for the training data, e.g. times series models and recurrent neural networks. The time dimension could also be taken into account by adding `memory' to the RFC so that predicted class of the storm is used as a feature for the prediction of following time step. 

So far, we also used only very basic methods for coping with missing data (just replace by zero) and imbalanced training data (using SMOTE). Imputing missing data with kriging interpolation based methods (first introduced in \cite{matheron1963principles}) would most probably improve the results. It would also be interesting to apply more advanced techniques for coping with imbalanced data, e.g. the  ``Rare-Transfer'' algorithm \cite{al2016transfer}.

Currently, Random Forest Classifier is used in an operational application.

\section*{Acknowledgements}

The authors would like to thank the project partners J\"ärvi-Suomen Energia, Loiste S\"ahkoverkko, and Imatra Seudun S\"ahk\"onsiirto for data and exprtice. Previous version of this work has been published as a conference paper [1].


\bibliographystyle{IEEEbib}
\bibliography{references}

\begin{thebibliography}{10}

\bibitem{Review2005Occurrence01}
Punkka Ari-Juhani and Bister Marja,
\newblock ``{Occurrence of Summertime Convective Precipitation and Mesoscale
  Convective Systems in Finland during 2000 – 01},''
\newblock {\em Monthly weather review}, vol. 133, no. 2, pp. 362--373, 2005.

\bibitem{galanaki2018thunderstorm}
E~Galanaki, K~Lagouvardos, V~Kotroni, E~Flaounas, and A~Argiriou,
\newblock ``Thunderstorm climatology in the mediterranean using cloud-to-ground
  lightning observations,''
\newblock {\em Atmospheric Research}, vol. 207, pp. 136--144, 2018.

\bibitem{Foote1979ResultsExperiment}
G~Brant Foote and Charles~A Knight,
\newblock ``{Results of a randomized hail suppression experiment in northeast
  Colorado. Part I: Design and conduct of the experiment},''
\newblock {\em Journal of Applied Meteorology}, vol. 18, no. 12, pp.
  1526--1537, 1979.

\bibitem{Oy2017KESKEYTYSTILASTOi}
Esa Niemel\"{a},
\newblock ``{KESKEYTYSTILASTO 2017 (i)},''
\newblock Tech. {R}ep. 2018-06-14 11:51:52.916, Energiateollisuus Ry,
  Etel\"{a}ranta 10, 00130 Helsinki, Finland, 2018.

\bibitem{Guikema2014PredictingPlanning}
Seth~David Guikema, Roshanak Nateghi, Steven~M. Quiring, Andrea Staid,
  Allison~C. Reilly, and Michael Gao,
\newblock ``{Predicting Hurricane Power Outages to Support Storm Response
  Planning},''
\newblock {\em IEEE Access}, vol. 2, pp. 1364--1373, 2014.

\bibitem{Guikema2010PrestormSystems}
Seth~D. Guikema, Steven~M. Quiring, and Seung~Ryong Han,
\newblock ``{Prestorm Estimation of Hurricane Damage to Electric Power
  Distribution Systems},''
\newblock {\em Risk Analysis}, vol. 30, no. 12, pp. 1744--1752, 2010.

\bibitem{Nateghi2014PowerModels}
Roshanak Nateghi, Seth Guikema, and Steven~M. Quiring,
\newblock ``{Power Outage Estimation for Tropical Cyclones: Improved Accuracy
  with Simpler Models},''
\newblock {\em Risk Analysis}, vol. 34, no. 6, pp. 1069--1078, 2014.

\bibitem{Han2009ImprovingModels}
Seung~Ryong Han, Seth~D. Guikema, and Steven~M. Quiring,
\newblock ``{Improving the predictive accuracy of hurricane power outage
  forecasts using generalized additive models},''
\newblock {\em Risk Analysis}, vol. 29, no. 10, pp. 1443--1453, 2009.

\bibitem{Wang2017ASystems}
Guang Wang, Tianhua Xu, Tao Tang, Tangming Yuan, and Haifeng Wang,
\newblock ``{A Bayesian network model for prediction of weather-related
  failures in railway turnout systems},''
\newblock {\em Expert Systems with Applications}, vol. 69, pp. 247--256, 2017.

\bibitem{allen2014application}
M~Allen, S~Fernandez, O~Omitaomu, and K~Walker,
\newblock ``Application of hybrid geo-spatially granular fragility curves to
  improve power outage predictions,''
\newblock {\em Journal of Geography \& Natural Disasters}, vol. 4, no. 2, pp.
  1--6, 2014.

\bibitem{chen2016fuzzy}
Po-Chen Chen and Mladen Kezunovic,
\newblock ``Fuzzy logic approach to predictive risk analysis in distribution
  outage management,''
\newblock {\em IEEE Transactions on Smart Grid}, vol. 7, no. 6, pp. 2827--2836,
  2016.

\bibitem{He2017NonparametricNetwork}
Jichao He, David~W. Wanik, Brian~M. Hartman, Emmanouil~N. Anagnostou, Marina
  Astitha, and Maria~E.B. Frediani,
\newblock ``{Nonparametric Tree-Based Predictive Modeling of Storm Outages on
  an Electric Distribution Network},''
\newblock {\em Risk Analysis}, vol. 37, no. 3, pp. 441--458, 2017.

\bibitem{Liu2018SceneNetwork}
Yanfei Liu, Yanfei Zhong, and Qianqing Qin,
\newblock ``{Scene Classification Based on Multiscale Convolutional Neural
  Network},''
\newblock {\em IEEE Transactions on Geoscience and Remote Sensing}, vol. 56,
  no. 12, pp. 7109 -- 7121, 7 2018.

\bibitem{li2015spatio}
Zhiguo Li, Amith Singhee, Haijing Wang, Abhishek Raman, Stuart Siegel,
  Fook-Luen Heng, Richard Mueller, and Gerard Labut,
\newblock ``Spatio-temporal forecasting of weather-driven damage in a
  distribution system,''
\newblock in {\em 2015 IEEE Power \& Energy Society General Meeting}. IEEE,
  2015, pp. 1--5.

\bibitem{singhee2017probabilistic}
Amith Singhee and Haijing Wang,
\newblock ``Probabilistic forecasts of service outage counts from severe
  weather in a distribution grid,''
\newblock in {\em 2017 IEEE Power \& Energy Society General Meeting}. IEEE,
  2017, pp. 1--5.

\bibitem{shield2018predictive}
Stephen Shield et~al.,
\newblock ``Predictive modeling of thunderstorm-related power outages,''
\newblock M.S. thesis, The Ohio State University, 2018.

\bibitem{Zhou2006ModelingLines}
Yujia Zhou, Anil Pahwa, and Shie~Shien Yang,
\newblock ``{Modeling weather-related failures of overhead distribution
  lines},''
\newblock {\em IEEE Transactions on Power Systems}, vol. 21, no. 4, pp.
  1683--1690, 2006.

\bibitem{kankanala2011regression}
P~Kankanala, A~Pahwa, and S~Das,
\newblock ``Regression models for outages due to wind and lightning on overhead
  distribution feeders,''
\newblock in {\em Power and Energy Society General Meeting, 2011 IEEE}. IEEE,
  2011, pp. 1--4.

\bibitem{Kankanala2012EstimationNetwork}
Padmavathy Kankanala, Anil Pahwa, and Sanjoy Das,
\newblock ``{Estimation of Overhead Distribution System Outages Caused by Wind
  and Lightning Using an Artificial Neural Network},''
\newblock in {\em International Conference on Power System Operation {\&}
  Planning}, 2012.

\bibitem{kankanala2014adaboost}
Padmavathy Kankanala, Sanjoy Das, and Anil Pahwa,
\newblock ``{AdaBoost\textsuperscript{+}: An Ensemble Learning Approach for
  Estimating Weather-Related Outages in Distribution Systems},''
\newblock {\em IEEE Transactions on Power Systems}, vol. 29, no. 1, pp.
  359--367, 2014.

\bibitem{Yue2018AData}
Meng Yue, Tami Toto, Michael~P. Jensen, Scott~E. Giangrande, and Robert Lofaro,
\newblock ``{A Bayesian approach-based outage prediction in electric utility
  systems using radar measurement data},''
\newblock {\em IEEE Transactions on Smart Grid}, vol. 9, no. 6, pp. 6149--6159,
  2018.

\bibitem{cintineo2014empirical}
John~L Cintineo, Michael~J Pavolonis, Justin~M Sieglaff, and Daniel~T Lindsey,
\newblock ``An empirical model for assessing the severe weather potential of
  developing convection,''
\newblock {\em Weather and Forecasting}, vol. 29, no. 3, pp. 639--653, 2014.

\bibitem{rossi2015object}
Pekka~Juhana Rossi,
\newblock {\em {Object-Oriented Analysis and Nowcasting of Convective Storms in
  Finland}},
\newblock Ph.D. thesis, Aalto University, 2015.

\bibitem{Makkonen2010SimulatingModel}
Lasse Makkonen and Bodo Wichura,
\newblock ``{Simulating wet snow loads on power line cables by a simple
  model},''
\newblock {\em Cold Regions Science and Technology}, vol. 61, no. 2-3, pp.
  73--81, 2010.

\bibitem{Rossi2013Real-timeData}
Pekka~J. Rossi, Vesa Hasu, Kalle Halmevaara, Antti~m?? Kel??, Jarmo Koistinen,
  and Heikki Pohjola,
\newblock ``{Real-time hazard approximation of long-lasting convective storms
  using emergency data},''
\newblock {\em Journal of Atmospheric and Oceanic Technology}, vol. 30, no. 3,
  pp. 538--555, 2013.

\bibitem{Dixon1993TITAN:Methodology}
Michael Dixon and Gerry Wiener,
\newblock ``{TITAN: Thunderstorm Identification, Tracking, Analysis, and
  Nowcasting—A Radar-based Methodology},''
\newblock {\em Journal of Atmospheric and Oceanic Technology}, vol. 10, no. 6,
  pp. 785--797, 1993.

\bibitem{Peura2002ComputerRemoval}
Markus Peura,
\newblock ``{Computer vision methods for anomaly removal},''
\newblock {\em Second European Conference on Radar Meteorology (ERAD02)}, pp.
  312--317, 2002.

\bibitem{sander1998density}
J{\"o}rg Sander, Martin Ester, Hans-Peter Kriegel, and Xiaowei Xu,
\newblock ``Density-based clustering in spatial databases: The algorithm
  gdbscan and its applications,''
\newblock {\em Data mining and knowledge discovery}, vol. 2, no. 2, pp.
  169--194, 1998.

\bibitem{ester1996density}
Martin Ester, Hans-Peter Kriegel, J\"{o}rg Sander, Xiaowei Xu, and {others},
\newblock ``{A density-based algorithm for discovering clusters in large
  spatial databases with noise.},''
\newblock in {\em KDD-96 Proceedings}, 1996, vol.~96, pp. 226--231.

\bibitem{Rossi2015KalmanStorms}
Pekka~J. Rossi, V.~Chandrasekar, Vesa Hasu, and Dmitri Moisseev,
\newblock ``{Kalman filtering-based probabilistic nowcasting of object-oriented
  tracked convective storms},''
\newblock {\em Journal of Atmospheric and Oceanic Technology}, vol. 32, no. 3,
  pp. 461--477, 2015.

\bibitem{Rossi2008AAnalysis}
Pekka Rossi and M\"{a}kel\"{a} Antti,
\newblock ``{A clustering-based tracking method for convective cell
  identification and analysis},''
\newblock {\em Fith European Conference on Radar in Meteorology and Hydrology},
  2008.

\bibitem{horn1981determining}
Berthold K~P Horn and Brian~G Schunck,
\newblock ``{Determining optical flow},''
\newblock {\em Artificial intelligence}, vol. 17, no. 1-3, pp. 185--203, 1981.

\bibitem{Bouguet2001PyramidalAlgorithm}
Jean-Yves Bouguet,
\newblock ``{Pyramidal implementation of the affine lucas kanade feature
  tracker description of the algorithm},''
\newblock {\em Intel Corporation}, vol. 5, no. 1-10, pp. 4, 2001.

\bibitem{Rossi2014AnalysisFinland}
Pekka~J Rossi, Vesa Hasu, Jarmo Koistinen, Dmitri Moisseev, Antti
  M{\"{a}}kel{\"{a}}, and Elena Saltikoff,
\newblock ``{Analysis of a statistically initialized fuzzy logic scheme for
  classifying the severity of convective storms in Finland},''
\newblock {\em Meteorological Applications}, vol. 21, no. 3, pp. 656--674,
  2014.

\bibitem{breiman2001random}
Leo Breiman,
\newblock ``{Random forests},''
\newblock {\em Machine learning}, vol. 45, no. 1, pp. 5--32, 2001.

\bibitem{Goodfellow-et-al-2016-mlp}
Ian Goodfellow, Yoshua Bengio, and Aaron Courville,
\newblock ``{Deep Learning},''
\newblock in {\em {Deep Learning}}, pp. 164--223. MIT Press, 2016.

\bibitem{Zhang2011AnData}
Xiuzhen Zhang and Yuxuan Li,
\newblock ``{An empirical study of learning from imbalanced data},''
\newblock {\em Conferences in Research and Practice in Information Technology
  Series}, vol. 115, pp. 85--94, 2011.

\bibitem{Pal2005RandomClassification}
M.~Pal,
\newblock ``{Random forest classifier for remote sensing classification},''
\newblock {\em International Journal of Remote Sensing}, vol. 26, no. 1, pp.
  217--222, 2005.

\bibitem{Goodfellow-et-al-2016-dropout}
Ian Goodfellow, Yoshua Bengio, and Aaron Courville,
\newblock ``{Deep Learning},''
\newblock in {\em {Deep Learning}}, pp. 255--265. MIT Press, 2016.

\bibitem{chawla2002smote}
Nitesh~V Chawla, Kevin~W Bowyer, Lawrence~O Hall, and W~Philip Kegelmeyer,
\newblock ``{SMOTE: synthetic minority over-sampling technique},''
\newblock {\em Journal of artificial intelligence research}, vol. 16, pp.
  321--357, 2002.

\bibitem{bergstra2012random}
James Bergstra and Yoshua Bengio,
\newblock ``Random search for hyper-parameter optimization,''
\newblock {\em Journal of Machine Learning Research}, vol. 13, no. Feb, pp.
  281--305, 2012.

\bibitem{kingma2014adam}
Diederik~P Kingma and Jimmy Ba,
\newblock ``{Adam: A Method for Stochastic Optimization},''
\newblock in {\em 3rd International Conference in Learning Representations},
  San Diego, 2015.

\bibitem{tsoumakas2009mining}
Grigorios Tsoumakas, Ioannis Katakis, and Ioannis Vlahavas,
\newblock ``Mining multi-label data,''
\newblock in {\em Data mining and knowledge discovery handbook}, pp. 667--685.
  Springer, 2009.

\bibitem{matheron1963principles}
Georges Matheron,
\newblock ``{Principles of geostatistics},''
\newblock {\em Economic geology}, vol. 58, no. 8, pp. 1246--1266, 1963.

\bibitem{al2016transfer}
Samir Al-Stouhi and Chandan~K Reddy,
\newblock ``{Transfer learning for class imbalance problems with inadequate
  data},''
\newblock {\em Knowledge and information systems}, vol. 48, no. 1, pp.
  201--228, 2016.

\end{thebibliography}

\begin{IEEEbiography}[{\includegraphics[width=1in,height=1.25in,clip,keepaspectratio]{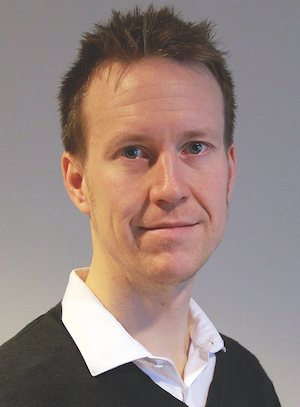}}]{Roope Tervo}
is a part-time Ph.D. student at Aalto University in Machine Learning Group with the main interest in impact analysis of the weather. The ultimate goal of his studies is to ennoble weather predictions to usable impact predictions. He also works as a Lead Architect at Finnish Meteorological Institute. 
\end{IEEEbiography}

\begin{IEEEbiography}[{\includegraphics[width=1in,height=1.25in,clip,keepaspectratio]{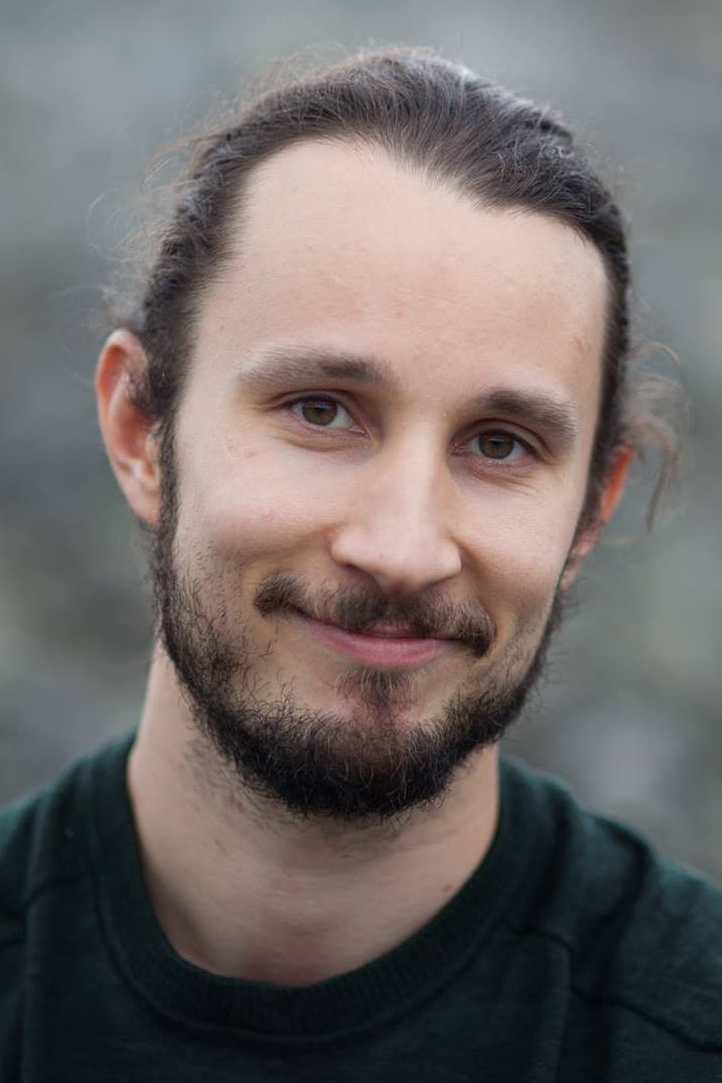}}]{Joonas Karjalainen}
Received M.Sc. degree in theoretical physics from the University of Oulu in 2013. Currently working as a group leader at the Finnish Meteorological Institute developing research-based applications and background systems for weather and marine data.
\end{IEEEbiography}

\begin{IEEEbiography}[{\includegraphics[width=1in,height=1.25in,clip,keepaspectratio]{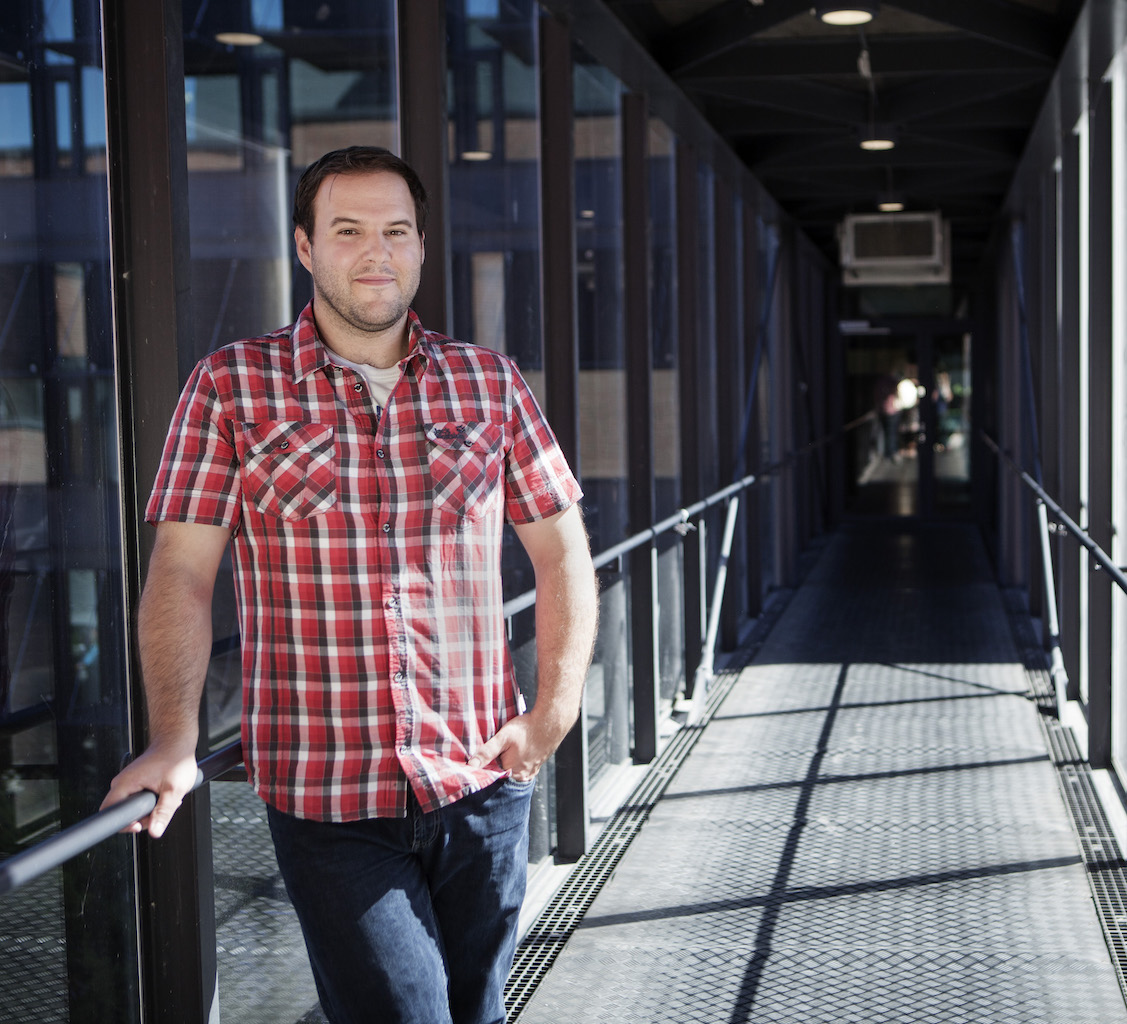}}]{Alexander Jung} is Assistant Professor at the Department of Computer Science of Aalto University. He obtained the  Master of Science (Tech.) and Ph.D. in Electrical Engineering and Computer Science from TU Vienna in 2008 and 2012, respectively. After a Post-Doc stay at ETH Zurich and TU Vienna, he joined Aalto in 2015 where he is leading the group “Machine Learning for Big Data”. His research revolves around fundamental limits and efficient algorithms for large-scale data analysis (big data). He received an Amazon Web Services Machine Learning award has been elected "Teacher of the Year" at the computer science department of Aalto University.
\end{IEEEbiography}

\end{document}